\documentclass[pre,aps,twocolumn,showpacs]{revtex4-2}
\usepackage{graphicx}
\usepackage{dcolumn}
\usepackage{bm}
\usepackage{color}
\usepackage{amssymb}
\usepackage{amsmath}
\usepackage [latin1]{inputenc}
\usepackage{tikz}
\usepackage{float}
\newcommand\VRule[1][\arrayrulewidth]{\vrule width #1}
\newcommand{\tikzcircle}[2][red,fill=red]{\tikz[baseline=-0.5ex]\draw[#1,radius=#2] (0,0) circle ;}

\begin{document}
\title{DNA Barcodes using a Cylindrical Nanopore}
\author{Swarnadeep Seth}
\author{Aniket Bhattacharya}
\altaffiliation[]
{Author to whom the correspondence should be addressed}
{}
\email{Aniket.Bhattacharya@ucf.edu}
\affiliation{$^1$Department of Physics, University of Central Florida, Orlando, Florida 32816-2385, USA}
\date{\today}
\begin{abstract}
We report an accurate method to determine DNA barcodes from the dwell time measurement of protein tags (barcodes) along the DNA backbone using Brownian dynamics simulation of a model DNA and use a recursive theoretical scheme which improves the measurements to almost 100\% accuracy. The heavier protein tags along the DNA backbone introduce a large speed variation in the chain that can be understood using the idea of non-equilibrium tension propagation theory. However, from an initial rough characterization of velocities into ``fast'' (nucleotides) and ``slow'' (protein tags) domains, we introduce a physically motivated interpolation scheme that enables us to determine the barcode velocities rather accurately. Our theoretical analysis of the motion of the DNA through a cylindrical nanopore opens up the possibility of its experimental realization and carries over to multi-nanopore devices used for barcoding.
\end{abstract}
\maketitle
A DNA barcode consists of a short strand of DNA sequence taken from a targeted gene like COI or cox I (Cytochrome C Oxidase 1)~\cite{barcode_CoxI} present in the mitochondrial gene in animals. The unique combination of nucleotide bases in barcode allows us to distinguish one species from another.
Unlike relying on the traditional taxonomical identification methods, DNA barcoding provides an alternative and reliable framework to categorize a wide variety of specimens obtained from the natural environment. Though researchers relied on DNA sequencing techniques for the identification of unknown species for a long time, in 2003, Hebert {\em et al.}~\cite{barcode_Hebert} proposed the mictocondrial gene (COI) region barcoding to classify cryptic species~\cite{barcode_cryptic_species} from the entire animal population. Since then, several studies have shown the potential applications of barcoding in conserving biodiversity~\cite{barcode_bio_diversity}, estimating phyletic diversity, identifying disease vectors~\cite{barcode_vector}, authenticating herbal products~\cite{barcode_medicinal_plant},  unambiguously labeling the food products~\cite{barcode_food_lebel,barcode_seafood}, and protecting endangered species~\cite{barcode_bio_diversity}. 
Traditional sequencing methods based on chemical analysis are widely used in the biological community to determine the barcodes.  Nanopore based sequencing methods~\cite{Dekker-2016} are being  explored in a dual nanopore system for a cost effective, high throughput, chemical-free, and real time barcode generation. 

\begin{figure}[ht!]
\includegraphics[width=0.47\textwidth]{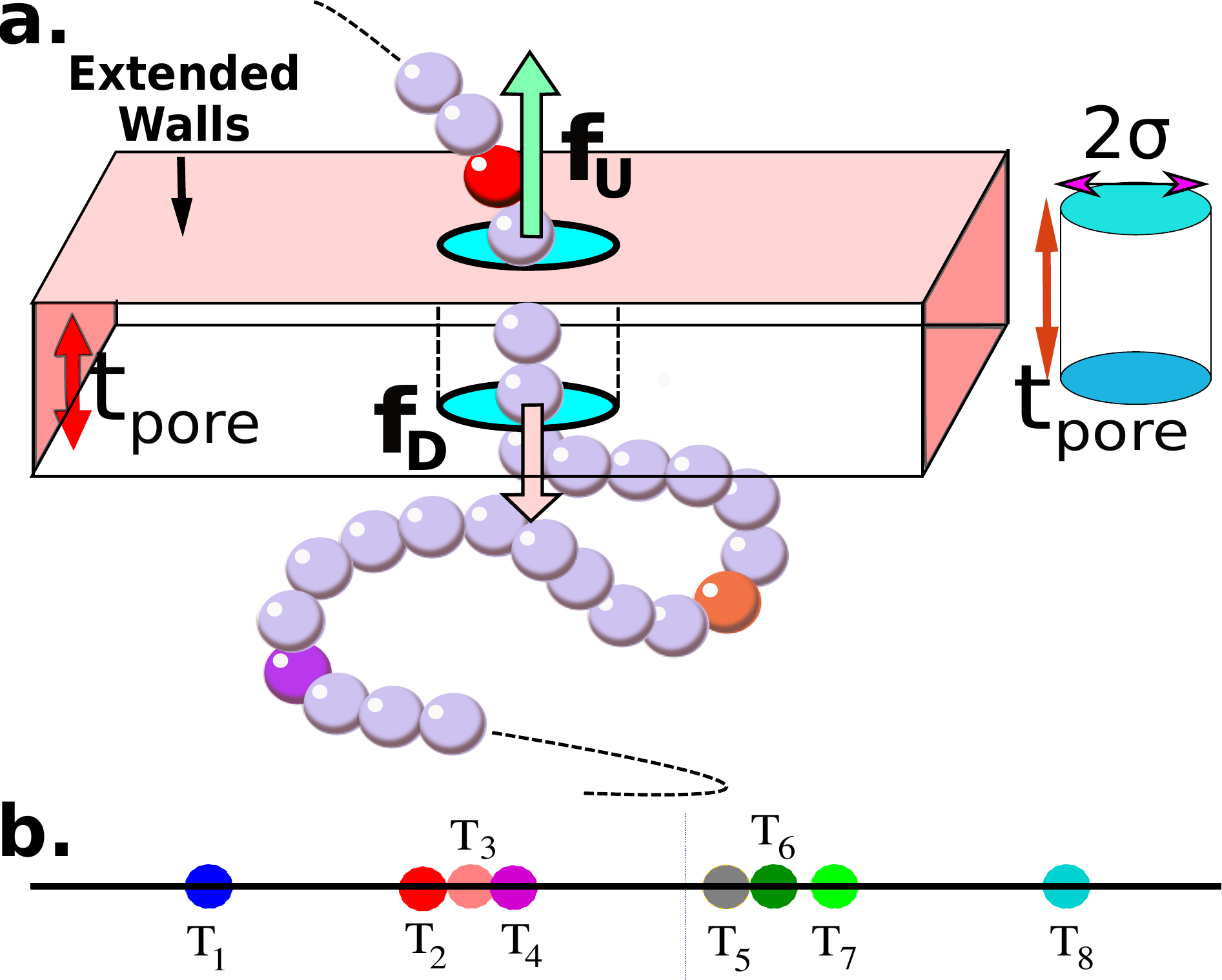}\\
\caption{\small \label{Model} Schematics of a model dsDNA captured in cylindrical  nanopore of diameter $d=2\sigma$ and  thickness $t_{pore}$, where $\sigma$ is the diameter of each monomer (purple beads). Protein tags (barcodes) of the same diameter but of different colors (only three are shown in here) interspersed along the dsDNA backbone. Opposite but unequal forces $\vec{f}_{U}$ and $\vec{f}_{D}$ are applied to straighten the dsDNA as it translocates in the direction bias net $\pm \vert \Delta \vec{f}_{UD}\vert =\pm \vert \vec{f}_{U}-\vec{f}_{D}\vert $ through the nanopore.
(b) Positions of the protein tags along the contour length of the model dsDNA of length $L=1024\sigma$ which represents an actual dsDNA of 48500 base pairs. The location of the tags are listed in Table-I.}
\end{figure}
\par
The possibility of determining DNA barcodes have been demonstrated in a dual nanopore device, by scanning a captured dsDNA multiple times by applying a net periodic bias across the two pores~\cite{Dekker-2016,Reisner-Small-2018,Reisner-Small-2019,Reisner-Small-2020}. Theoretical and simulation studies have also been reported in the context of a double nanopore
system~\cite{Bhattacharya_Seth_2020,Seth-JCP-2020, Aksimentiev-2020}. In this article, we investigate a similar strategy {\em in silico} in a cylindrical nanopore and demonstrate that a cylindrical nanopore can have a competitive advantage over a dual nanopore system. By studying a  model dsDNA with barcodes using Brownian dynamics we establish an important result that it is due to the disparate dwell time and speed of the barcodes (``tags'') compared to the nucleotide segments
\begin{table}[ht!]
\caption{Tag positions along the dsDNA}
\resizebox{\columnwidth}{!}{\begin{tabular}{l l l l l l l l l}\hline  
Tag \# & \color{blue} \bf{}$T_1$ ~~& \color{red} \bf{}$T_2$ ~~& \color{orange} \bf{}$T_3$ ~~& \color{magenta} \bf{}$T_4$ ~~& \color{darkgray} \bf{}$T_5$ ~~& \color{teal} \bf{}$T_6$ ~~& \color{green} \bf{}$T_7$ ~~& \color{cyan} \bf{}$T_8$\\ \hline  
Position & 154 & 369 & 379 & 399 & 614 & 625 & 696 & 901 \\ \hline  
Separation &  154 & 215 & 10 & 20 & 215 & 11 & 71 & 205 \\ \hline 
\end{tabular}
}
\end{table}
(``monomers'') the current blockade time information only is not enough and will lead to an inevitable underestimation of the distance between the barcodes. Furthermore, using the ideas of the tension propagation theory~\cite{Sakaue_PRE_2007,Ikonen_JCP2012}, we demonstrate that information about  the fast-moving nucleotides in between the barcodes,- not easily accessible experimentally is a key element to resolve the underestimation. We suggest how to obtain this information experimentally and provide a  physically motivated ``two-step'' interpolation scheme for an accurate determination of barcodes, even when the separation of (unknown) tags has a broad distribution.
\par
\vskip 0.5truecm
$\bullet$~{\em The Model System:}~Our {\em in silico} coarse-grained (CG) model of a dsDNA consist of 1024 monomers interspersed with 8 barcodes at different locations  shown in Fig.~\ref{Model} and Table-I is motivated by an experimental study by Zhang {\em et al.} on a 48500 bp long dsDNA with 75 bp long protein tags at random locations along the chain ~\cite{Reisner-Small-2018, Reisner-Small-2019,Reisner-Small-2020} using a dual nanopore device. Here we explore if a cylindrical nanopore with applied biases at each end can resolve the barcodes with similar accuracy or better.  We purposely choose positions of the 8 barcodes (Table-I) to study how the effect of disparate distances among the barcodes affects their measurements.
The tags {\bf \color{red} $T_2$}, {\bf \color{orange} $T_3$},  {\bf \color{magenta} $T_4$} are closely spaced and form a group. Likewise, another group consisting of {\bf\color{darkgray} $T_5$} and {\bf\color{teal} $T_6$} are put in a closer proximity to  {\bf\color{green} $T_7$}. The tags {\bf \color{blue} $T_1$ } and {\bf \color{cyan} $T_8$ } are further apart from the rest of the tags.
The general scheme of the BD simulation strategy for a translocating homo-polymer under alternate bias has been discussed in our recent publication~\cite{Bhattacharya_Seth_2020,Seth-JCP-2020} and in the Appendix~A. 
\par
\begin{figure}[ht!]
\includegraphics[width=0.45\textwidth]{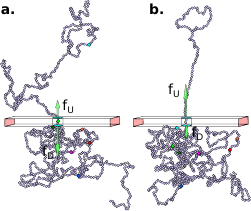}
\caption{\small Demonstration of the epoch when the bias voltage is flipped. (a) showing the last barcode is yet to translocate in the downward direction when the net bias $\Delta \vec{f}_{DU} = \vec{f}_D - \vec{f}_U > 0$. (b) shows the situation after a later time when finally all the barcodes crossed the cylindrical pore during downward translocation with a portion of the end segment  still remaining inside the pore.  At this point the bias is flipped with an upward bias
$\Delta \vec{f}_{UD} = \vec{f}_U - \vec{f}_D > 0$, translocation now occurs in the upward direction.  In this way, the DNA remains captured all the time during repeated scans.}
\label{flip}
\end{figure}

In this article, tags are introduced by choosing the mass and friction coefficient at tag locations to be
different than the rest of the monomers along the chain. This 
requires modification of the BD algorithm as discussed in the
Appendix~A.  The protein tags used in the experiments 
\cite{Reisner-Small-2018,Reisner-Small-2019,Reisner-Small-2020} translate to about three monomers in the simulation. The heavier and extended tags introduce a larger viscous drag. Instead of explicitly putting side-chains at the tag locations, we made the mass and the friction coefficient of the tags 3 times larger. This we find enough to resolve the distance between the tags. Two forces $\vec{f}_U$ and $\vec{f}_D$ at each end of the cylinder in opposite directions keep the DNA straight inside the channel and allows translocation in the direction of the net bias (please see Fig.~\ref{Model} and Fig.~\ref{flip}). 
\par
\vskip 0.5truecm
$\bullet$~{\em Barcodes from repeated scanning:} As potentially could be done in a nanopore experiments, we switch the differential bias once the first tag or the last tag
({\bf \color{blue} $T_1$ }, {\bf \color{teal} $T_8$}) translocates through the nanopore during up($U$)/down($D$) $\rightarrow D/U$ translocation yet having end segments inside the
pore (please see Fig.~\ref{flip}) so that  the DNA remains captured in the cylindrical pore and the barcodes are scanned multiple times.
\begin{figure}[ht!]
\includegraphics[width=0.43\textwidth]{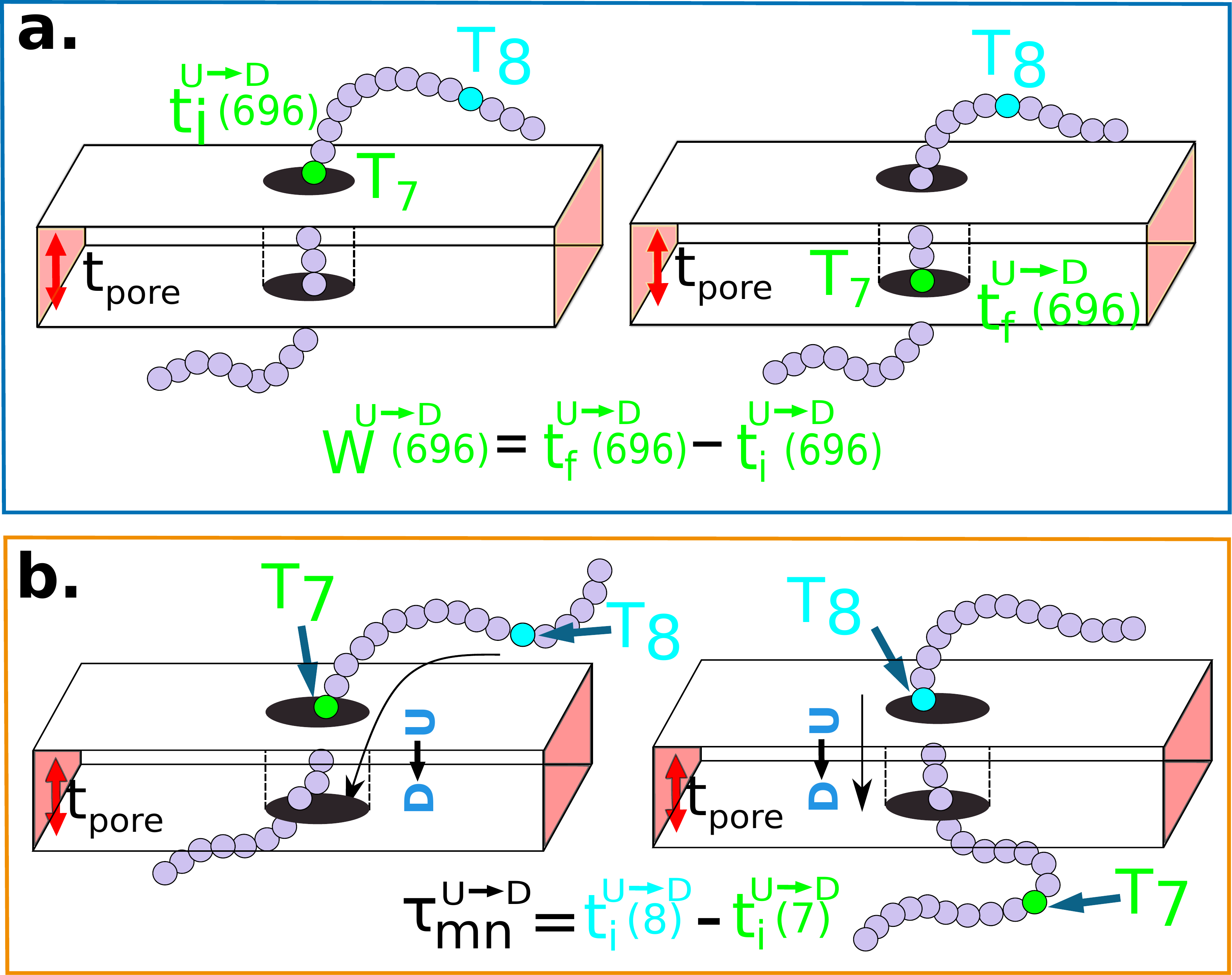}
\caption{\small (a) Demonstration of calculation of wait time for  {\bf\color{green} $T_7$} which has the monomer of index $m=696$. The dwell velocity is then calculated using Eqn.~\ref{v_dwell}. 
(b) Demonstration of calculation of tag time delay $\tau_{78}^{U \rightarrow D}= t_i^{U \rightarrow D}(8) - t_i^{U \rightarrow D}(7)$ for  tags {\bf\color{green} $T_7$}  and {\bf\color{cyan} $T_8$} while they are moving downward. Please note that similar quantity for upward translocation  $\tau_{87}^{D \rightarrow U}= t_i^{D \rightarrow U}(7) - t_i^{U \rightarrow D}(8)   \ne \tau_{78}^{U \rightarrow D} $ as there is no symmetry of the tags along the chain.}
\label{dwell_demo}
\end{figure}
The question we ask: can we recover the actual barcode locations  from these scanning measurements, so that the method can be applied to determine unknown barcodes ? We monitor two important quantities, -  the dwell time of each monomer and the time delay of arrival of two successive monomers at the pore as demonstrated in Fig.~\ref{dwell_demo} and explained below. 
For each up/down-ward scan we measure the dwell times of the monomer $m$ as follows:
\begin{subequations}
\begin{gather}
W^{U \rightarrow D}(m) =t_f^{U \rightarrow D}(m) - t_i^{U \rightarrow D}(m), \\
W^{D \rightarrow U}(m)=t_f^{D \rightarrow U}(m) - t_i^{D \rightarrow U}(m).
\end{gather}
\label{t_dwell}
\end{subequations}
 Here $t_i^{U \rightarrow D}(m)$ and $t_f^{U \rightarrow D}(m)$ are the arrival and exit times of the monomer with index $m$ as further demonstrated in Fig.~\ref{dwell_demo}(a).
The corresponding dwell velocities $v_{dwell}^{U \rightarrow D}(m)$ and  $v_{dwell}^{D \rightarrow U}(m)$ for the $m^{th}$ bead (either a monomer or a tag) along the channel axis (please see  Fig.~\ref{dwell_demo}(a)) can be obtained as follows.
\begin{subequations}
  \begin{gather}
v_{dwell}^{U \rightarrow D}(m) = t_{pore}/W^{U \rightarrow D}(m), \\
v_{dwell}^{D \rightarrow U}(m) = t_{pore}/W^{D \rightarrow U}(m).
\end{gather}
\label{v_dwell}
\end{subequations}
In an actual experiment one measures the dwell velocities of the tags only which are equivalent to the current blockade times. \par
\begin{figure}[ht!]
\includegraphics[width=0.45\textwidth]{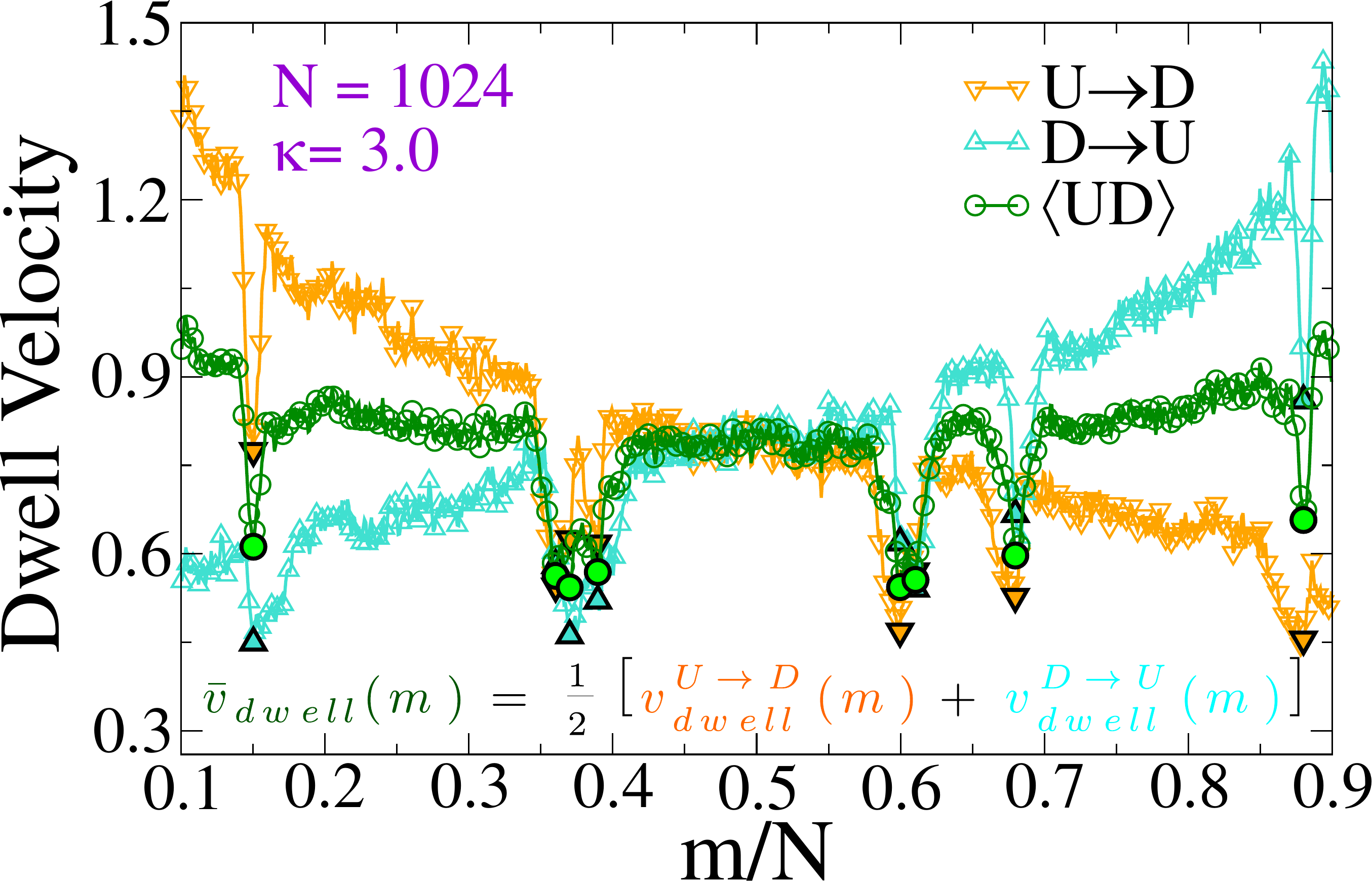}
\caption{\small Dwell velocity of monomer in a cylindrical nanopore system. {\color{orange}$\triangledown$}
and  {\color{cyan}$\vartriangle$} represent downward and upward translocation.  {\color{green}\large$\circ$}  are average of both directions. Filled triangles and circles correspond to tag dwell velocities.}
\label{dwell_vel}
\end{figure}

{\em $\bullet$~Non uniformity of the dwell velocity:} The presence of tags with heavier mass ($m_{tag} =3 m_{bulk}$) and larger solvent friction ($\gamma_{tag} = 3 \gamma_{bulk}$) introduces a large variation in the dwell time and hence a large variation in the dwell velocities of the DNA beads and tags (see Fig.~\ref{dwell_vel}). In general, there is no up-down symmetry for the dwell time/velocity as tags are not located symmetrically along the chain backbone.
Thus the physical quantities are averaged over $U \rightarrow D$ and
$D \rightarrow U$ translocation data. The average dwell velocity $\bar{v}_{dwell}(m) = \frac{1}{2} \left[v_{dwell}^{U \rightarrow D}(m)+v_{dwell}^{D \rightarrow U}(m) \right]$
clearly shows two different velocity envelopes - the tags residing at the lower envelope. Fig.~\ref{dwell_vel} shows that
the dwell velocities of the tags (green circle \tikzcircle[fill=green]{2.5pt}) are significantly lower than the velocity of the nucleotides in between the tags, which will underestimate the barcode distances as explained later. We further notice that  increasing the pore width resolves the barcodes better. \par
\begin{figure}[ht!]
\includegraphics[width=0.45\textwidth]{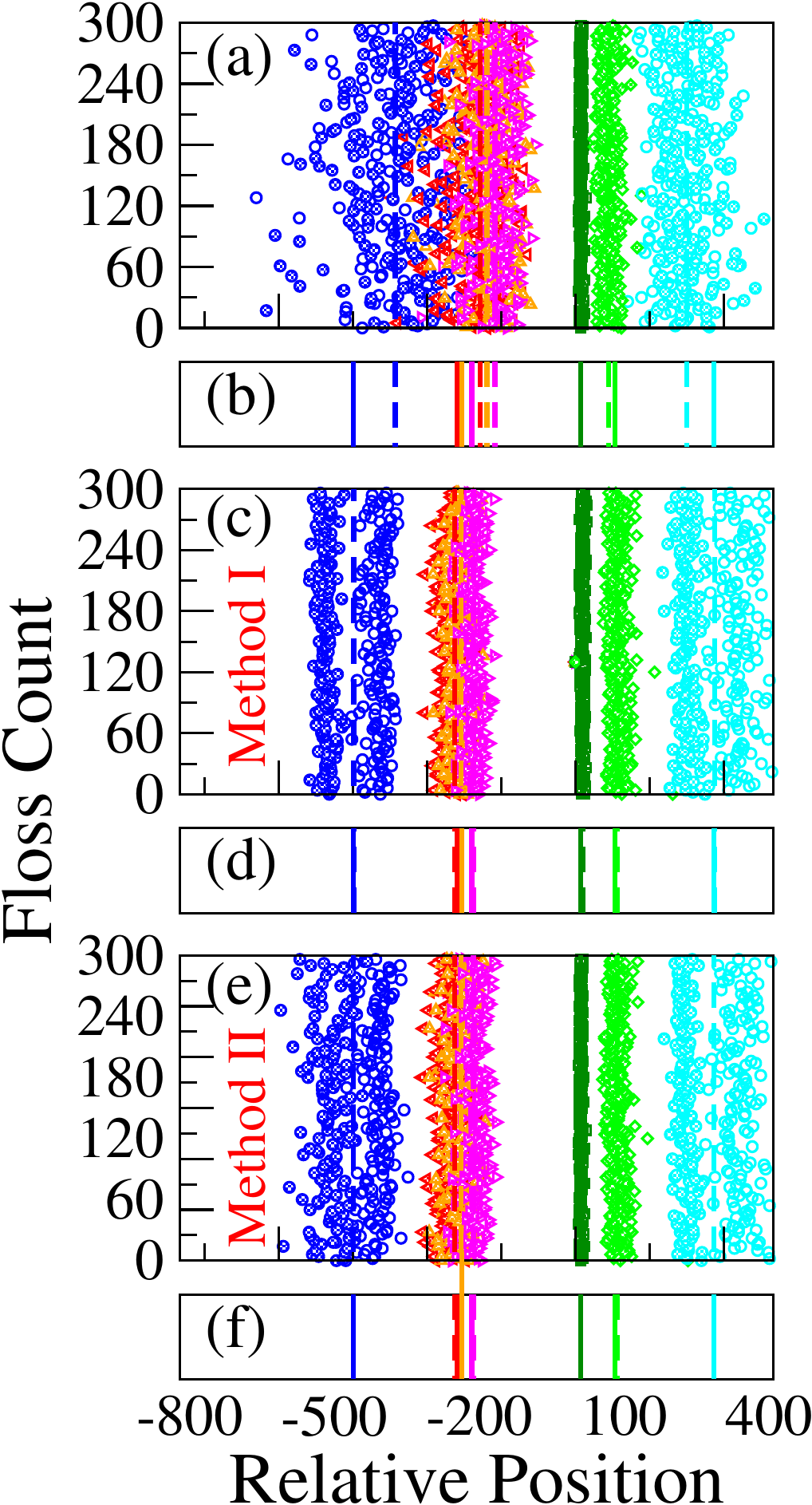}
\par 
\caption{\small \label{barcodes}(a) Barcodes generated using different methods. In each graph, 
the colored symbols/lines refer to the corresponding colors of the barcodes 
{\bf \color{blue} $T_1$}, {\bf \color{red} $T_2$ }, {\bf \color{orange} $T_3$}, {\bf \color{magenta}, $T_4$}, {\bf \color{darkgray} $T_5$}, {\bf \color{teal} $T_6$},
{\bf \color{green} $T_7$}, and {\bf \color{cyan} $T_8$} respectively. The open and filled symbols represent barcodes for  $U\rightarrow D$ and $D \rightarrow U$ transolcation using (a) Eqn. 3; (c) using method 1, and (e) using method 2. In (b), (d) and (e) the solid lines refer to the actual location of the barcodes and the dashed lines correspond to the averages from (a), (c) and (e) respectively. The improved accuracy for the latter two methods are readily visible in (d) and (f) where the simulation and the actual data are almost indistinguishable.}
\end{figure}

 
\vskip 0.2truecm
{\em $\bullet$ Barcode estimation using a cylindrical nanopore setup:}~If the dsDNA with barcodes were a rigid rod, then 
one could obtain the barcode distances $d_{mn}^{U \rightarrow D}$  and $d_{nm}^{D \rightarrow U}$  between tags $T_m$ and $T_n$  from the following equations (shown for downward translocation only):
\begin{subequations} \label{short}
\begin{align}
  d_{mn}^{U \rightarrow D}& = v_{mn}^{U \rightarrow D} \times \tau_{mn}^ {U \rightarrow D} \quad{\rm where,}\\
  v_{mn}^{U \rightarrow D} & = \frac{1}{2} \left[  v_{dwell}^{U \rightarrow D}(m)  + v_{dwell}^{U \rightarrow D}(n)  \right],\\
   \tau_{mn}^ {U \rightarrow D} & =  \left( t_i^{U \rightarrow D}(n) - t_i^{U \rightarrow D}(m) \right ) . 
 \end{align}
\end{subequations}
Here $\tau_{mn}^ {U \rightarrow D}$  is the time delay of arrivals of $T_m$ and $T_n$ for downward translocation (please see Fig.~\ref{dwell_demo}(b) which explains the special case when $m=7$ and $n=8$).  Similar Equations can be obtained by flipping $D$ and $m$ with $U$ and $n$ respectively.  In other words, Eqn.~\ref{short} gives the shortest distance and not necessarily the contour length (the actual distance) between the tags. However, this is the only data accessible through experiments and likely to provide an underestimation of the barcodes. Fig.~\ref{barcodes}(a) shows the data for 300 scans. The average with error bars are shown in the 3$^{rd}$ column of Table-II. Excepting for {\bf\color{teal} $T_6$}  these measurements grossly underestimate the actual positions with large error bars.\par
\par
\begin{table}[ht!]
\caption{Barcodes from various methods}
\centering
\begin{tabular}{!{\color{violet}\VRule[1pt]} c !{\color{green}\VRule[1pt]} c !{\color{red}\VRule[1pt]} l !{\color{red}\VRule[1pt]} l !{\color{violet}\VRule[1pt]}  l !{\color{violet}\VRule[1pt]}} \hline 
Tag & Relative  & Barcode & Barcode & Barcode  \\
Label & Distance & (Eqn.~\ref{short}) &(Method-I) & (Method-II) \\
& w.r.t $T_5$ & ~~\color{red} \bf{$\times$}  & ~~\color{green} \bf{$\checkmark$} &~~ \color{green} \bf{$\checkmark$} \\ \hline
\color{blue} \bf{}$T_1$  & 460  & 373 $\pm$ 122 & 459 $\pm$ 59 & 460 $\pm$ 43 \\ \hline 
\color{red} \bf{}$T_2$ & 245 & 197 $\pm$ 67 & 250 $\pm$ 39 & 250 $\pm$ 32 \\ \hline 
\color{orange} \bf{}$T_3$ & 235 & 183 $\pm$ 63 & 237 $\pm$ 38 & 237 $\pm$ 32 \\ \hline 
\color{magenta} \bf{}$T_4$ & 215 & 167 $\pm$ 54  & 211 $\pm$ 35 & 211 $\pm$ 30 \\ \hline 
\color{darkgray} \bf{}$T_5$ & 0 & 0 & 0 & 0 \\ \hline 
\color{teal} \bf{}$T_6$ & 11 & 11 $\pm$ 3 & 14 $\pm$ 4 & 11 $\pm$ 3 \\ \hline 
\color{green} \bf{}$T_7$  & 82 & 68 $\pm$ 23 & 86 $\pm$ 23 & 86 $\pm$ 21 \\ \hline 
\color{cyan} \bf{}$T_8$ & 287 & 230 $\pm$ 73 & 287 $\pm$ 65 & 287 $\pm$ 73 \\ \hline 
\end{tabular}
\end{table}
\par
\begin{figure}[ht!]
\includegraphics[width=0.45\textwidth]{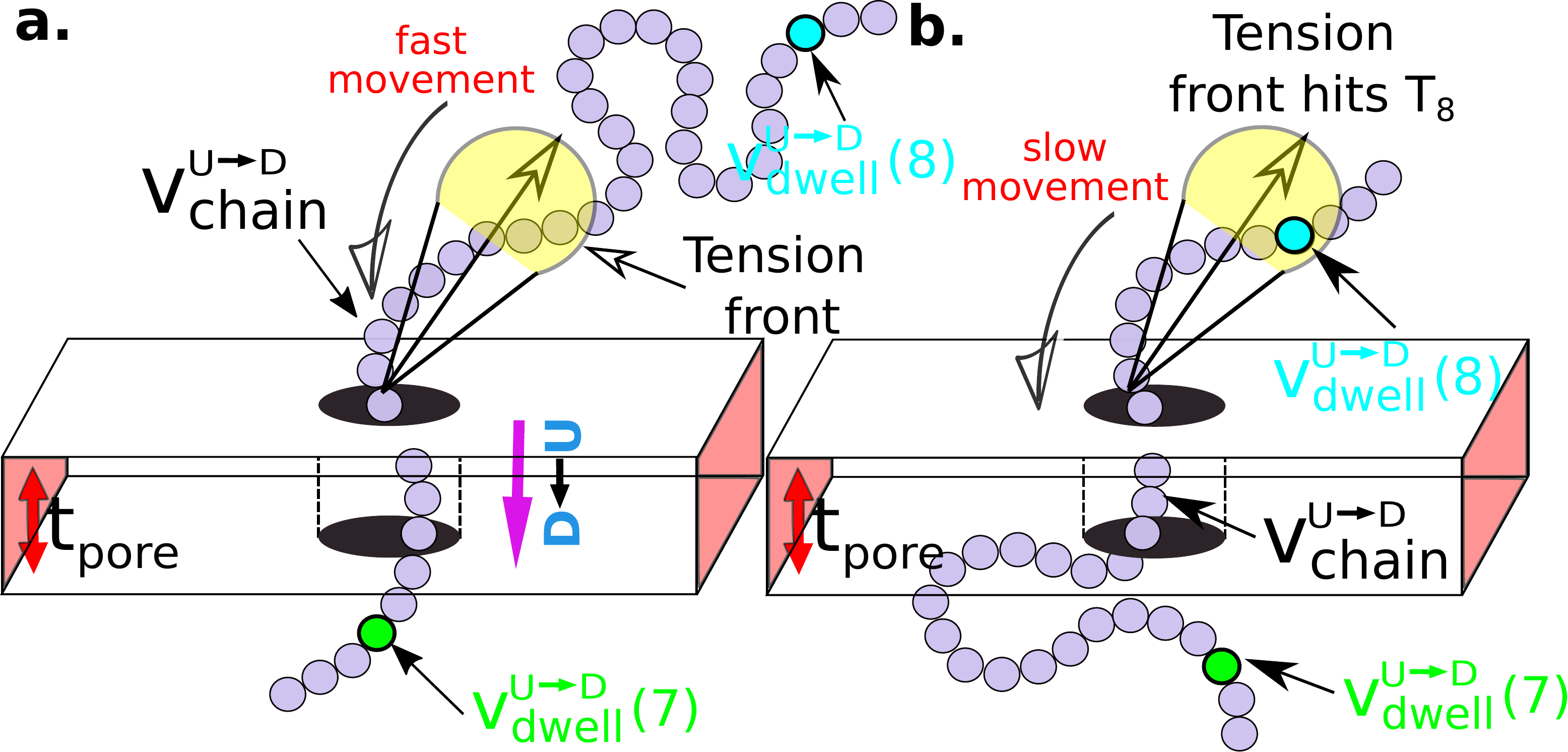}
\caption{\small \label{TP} Tension propagation (TP) through the chain backbone connecting {\bf \color{green}$T_7$} and {\bf \color{cyan} $T8$}. (a) Figure shows a sudden fast movement of monomers right after {\bf \color{green}$T_7$}'s passage through the pore. Due to the TP front's influence (yellow blob region), subsequent monomers are sucked into the pore quickly. (b) TP front finally reaches {\bf \color{cyan} $T8$}, leading to a slower translocation speed due to the tag's large inertia and higher viscous drag.}
\end{figure}

\vskip 0.2truecm
{\em $\bullet$~Tension Propagation (TP) Theory explains the source of discrepancy and provides solution:}~
Unlike a rigid rod, tension propagation governs the semi-flexible chain's motion in the presence of an external bias. 
In TP theory and its implementation  in Brownian dynamics, the motion of the subchain in the {\em cis} side decouples into two domains~\cite{Sakaue_PRE_2007, Ikonen_JCP2012}. In the vicinity of the pore, the tension front affects the motion directly while the second domain remains unperturbed, beyond the reach of the TP front. In our case, after the tag $T_m$ translocates through the pore, preceding monomers are dragged into the pore quickly by the tension front, analogous to the uncoiling effect of a rope pulled from one end. 
The onset of this sudden {\em faster} motion continues to grow and reaches its maximum until the tension front hits the subsequent tag $T_{m \pm1}$, with larger inertia and viscous drag. At this time (called the tension propagation time~\cite{Adhikari_JCP_2013})  the faster motion of the monomers begins to taper down to the velocity of the tag $T_{m \pm1}$. This process continues from one segment to the other. Fig.~\ref{TP} shows an example on how the segment connecting  {\bf \color{green}$T_7$} and {\bf \color{cyan} $T8$} has non-monotonic velocity under the influence of the tension front. 
 These contour lengths of faster moving segments in between two barcodes are not  accounted for in Eqn~\ref{short}. The experimental protocols are limited in extracting barcode information through Eqn.~\ref{short} (measuring current blockade time) and therefore, likely to underestimate the barcodes, unless the data is corrected to account for the faster moving monomers in between two tags.  
\par
\vskip 0.2truecm
{\em $\bullet$ How to determine the barcodes correctly ?}~
Fig.~\ref{Model}(b) and the $3^{rd}$ column of Table-II when looked closely provide clues to the solution of the underestimated tag distances. We note that
locations of the isolated tags (such as,  {\bf \color{blue} $T_1$} and   {\bf \color{cyan} $T_8$}) far from {\bf \color{darkgray} $T_5$} have a larger error bar while {\bf \color{teal} $T_6$} which is adjacent to {\bf \color{darkgray} $T_5$}  has the correct distance from Eqn.~\ref{short}. It is simply because in the later case the contour length between
{\bf \color{darkgray} $T_5$} and {\bf \color{teal} $T_6$} is almost equal to the shortest distance. Evidently, the error bars increase with increased separation.
\par
\begin{figure}[ht!]
\includegraphics[width=0.47\textwidth]{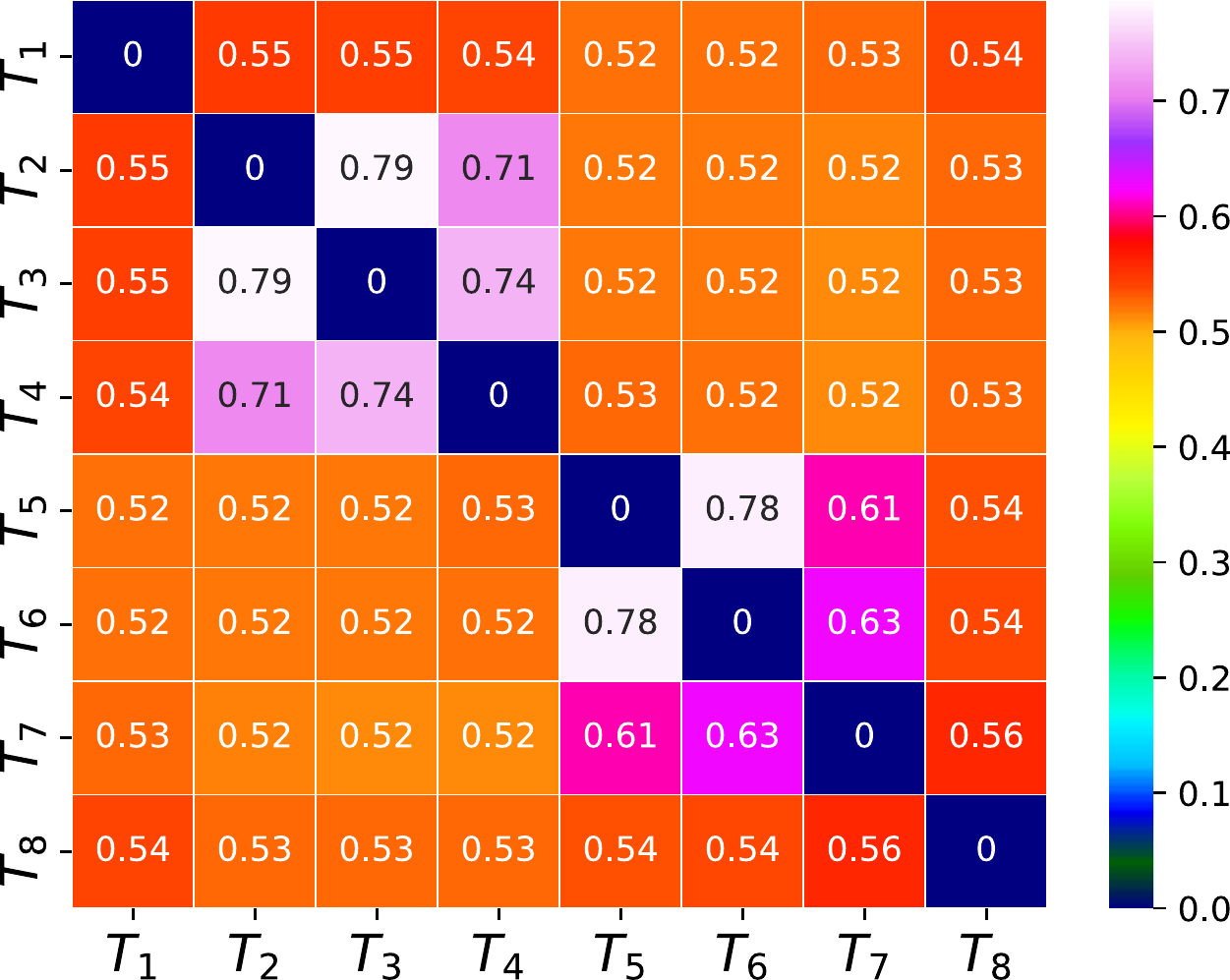}
\caption{\small \label{heatmap} Flory exponent ($\nu$) for the segment connecting a tag pair represented as a two dimensional heatmap array on the color scale ranging from blue to white.}
\end{figure}

To compare the barcodes obtained from Eqn.~\ref{short} with the actual contour length (see $2^{nd}$ column of Table-II) between tag pairs, we invoke the Flory theory to determine the scaling exponent $\nu$ ~\cite{Rubinstein} which reveals the behavior of the segments under translocation. The heatmap in Fig.~\ref{heatmap} confirms that when the separation between the tag pairs is less compared to the DNA length, the connecting segment behaves like a rigid rod ($\nu > 0.6$). While for the isolated tags, $\nu < 0.6$ suggests that barcodes are shorter than their respective contour lengths. This clarifies the reason behind the barcode underestimation for the tags which are spaced apart while yielding accurate barcodes for tags located in groups.

\par
Within the experimental set up we suggest the following two methods which will account for the larger velocities of the monomers. 
\par
\vskip 0.2truecm
{\em Method 1 - Barcode from known end-to-end Tag distance:}~
In order to measure the barcode distances accurately one thus needs the velocity of the entire chain. If the distance between {\bf \color{blue} $T_1$} 
and {\bf \color{cyan} $T_8$}) $d_{18}\simeq L $, then the velocity of the segment $d_{18}$  will approximately account for the average velocity of the entire chain $v_{chain}$ and correct the problem as demonstrated next. First we estimate the velocity of the chain
\begin{equation}
v_{chain}^{U \rightarrow D} \approx v_{18}^{U \rightarrow D} = d_{18}/\tau_{18}^{U \rightarrow D},
\label{v_chain}
\end{equation}
assuming we know $d_{18}$ and 
$\tau_{18}^{U \rightarrow D}$  is the time delay of arrival at the pore between {\bf \color{blue} $T_1$} and {\bf \color{cyan} $T_8$} for ${U \rightarrow D}$ translocation. We then estimate the barcode distance  $d_{mn}^{U \rightarrow D}$ between tags $T_m$ and $T_n$ as 
\begin{equation}
d_{mn}^{U \rightarrow D} =  v_{18}^{U \rightarrow D} \times \tau_{mn}^{U \rightarrow D}.
\label{method1}
\end{equation}
In the similar fashion one can calculate $d_{mn}^{D \rightarrow U}$ using $v_{chain}^{D \rightarrow U}$  and $\tau_{mn}^{U \rightarrow D}$ information respectively. 
How do we know $d_{18}$ ? One can use  $d_{18} \approx  L_{\rm scan}$ and  $v_{chain} \approx \bar{v}_{\rm scan}$,  from Eqn.~\ref{vscan} where $\bar{v}_{\rm scan}$ is the the average velocity of the scanned length $L_{\rm scan}$ from repeated scanning as discussed in the next paragraph. 
This method is effective for estimating the long-spaced barcodes but it overestimates the barcode distance if multiple barcodes are close by as evident in Fig.~\ref{barcodes}(d) and the $4^{th}$ column of table-II. Thus, we know how to obtain barcode distances accurately when they are close by  (from Eqn.~\ref{short}) and for large separation (Eqn.~\ref{method1}). We now apply the physics behind these two schemes to derive an interpolation scheme that will work for all separations among the barcodes.
\par
{\em Method 2 - Barcode using two-step method:}~Average scan time $\bar{\tau}_{\rm scan}$ for the entire chain (which can be measured experimentally)  is a better way to estimate the average velocity of the chain. $L_{\rm scan}$ is the maximum length up to which the dsDNA segment remains captured inside the nanopore gets scanned and denotes the theoretical maximum beyond which the dsDNA will escape from the nanopore, thus, $L \approx L_{\rm scan}$. For example, in our simulation, scanning length $L_{\rm scan}=0.804L$. We denote the average scan velocity as
\begin{equation}
\bar{v}_{\rm scan} = \frac{1}{N_{\rm scan}}\sum_{i=1}^{N_{\rm scan}} L_{\rm scan}/\tau_{\rm scan}(i),
\label{vscan}
\end{equation}
where $\tau_{\rm scan}(i)$ is the scan time for the $i^{th}$ event, and $N_{\rm scan}=300$. 
To proceed further, we use our established results that the monomers of the dsDNA segments in between the tags move with velocity $\bar{v}_{\rm scan}$, while tags move with their respective dwell velocities $v_{mn}^{U \rightarrow D}$ and $v_{mn}^{D \rightarrow U}$ (Eqn.~\ref{v_dwell}). We then calculate the segment velocity between two tags by taking the {\em weighted average} of the velocities of tags and DNA segment in between as follows.
\par
First, we estimate the approximate number of monomers $N_{mn}=d_{mn}^{U \rightarrow D}/\langle b_l \rangle $
($\langle b_l \rangle$  is the bond-length)  by considering the tag velocities only using Eqn.~\ref{short}.  We then calculate the segment velocity accurately by incorporating weighted velocity contributions from both the tags and the monomers between the tags.
\begin{equation}
\begin{split}
v_{weight}^{U \rightarrow D} = \frac{1}{N_{mn}} \Big[ v_{dwell}^{U \rightarrow D}(m)  &+  v_{dwell}^{U \rightarrow D}(n) + \\
& (N_{mn}-2) \bar{v}_{\rm scan} \Big]
\end{split}
\label{twostep}
\end{equation}
The barcodes are finally estimated by multiplying the calculated 2-step velocity in Eqn.~\ref{twostep} above by the tag time delay as
\begin{equation}
d_{mn}^{U \rightarrow D} = v_{weight}^{U \rightarrow D}\times \tau_{mn}^{U \rightarrow D}
\end{equation}
for $U \rightarrow D$ translocation and repeating the procedure for $D \rightarrow U$ translocation. 
This 2-step method accurately captures the distance between the barcodes when the two tags are in proximity or spaced apart from each other. Table-II and Fig.~\ref{barcodes} summarize our main results and claims.
\par
\vskip 0.2truecm
{\em $\bullet$~Summary \& Future work:}~
Motivated by the recent experiments we have designed barcode determination experiment {\em in silico} in a cylindrical nanopore using the Brownian dynamics scheme on a model dsDNA with known locations of the barcodes. We have carefully chosen the locations of the barcodes so that the separations among the barcodes span a broad distribution. We discover that if we use the dwell time data only for the barcodes from multiple scans of the dsDNA to calculate the average velocities of the tags then the method underscores the barcode distances for tags further apart. Our simulation guides us to conclude that the source of this underestimation lies in neglecting the information contained in the faster moving DNA segments in between any two tags. We use non-equilibrium tension propagation theory to explain the non-monotonic velocity of the chain segments where the barcodes lie at the lower bound of the velocity envelope as shown in Fig.~\ref{dwell_vel}. The emerging picture readily shows the way how to rectify this error by introducing an interpolation scheme that works well to determine barcodes spaced apart for all distances which we validate using simulation data. We suggest how to implement the scheme in an experimental setup. It is important to note that the interpolation scheme-based concept of the TP theory is quite general and we have ample evidence that this will work in a double nanopore system as well.
\par
\vskip 0.2truecm
{\em $\bullet$~Conflicts of interest:}
The authors declare no competing financial interest.
\par
\vskip 0.2truecm
{\em $\bullet$~Acknowledgements:}
The research at UCF has been supported by the grant number 1R21HG011236-01 from the National Human Genome Research Institute at the National Institute of Health. All computations were carried out at the UCF's high performance computing platform STOKES.

\appendix
\setcounter{figure}{0}
\renewcommand{\thefigure}{A\arabic{figure}}
\section{The Model and Brownian dynamics simulation} 
\begin{figure}[ht!]
\includegraphics[width=0.47\textwidth]{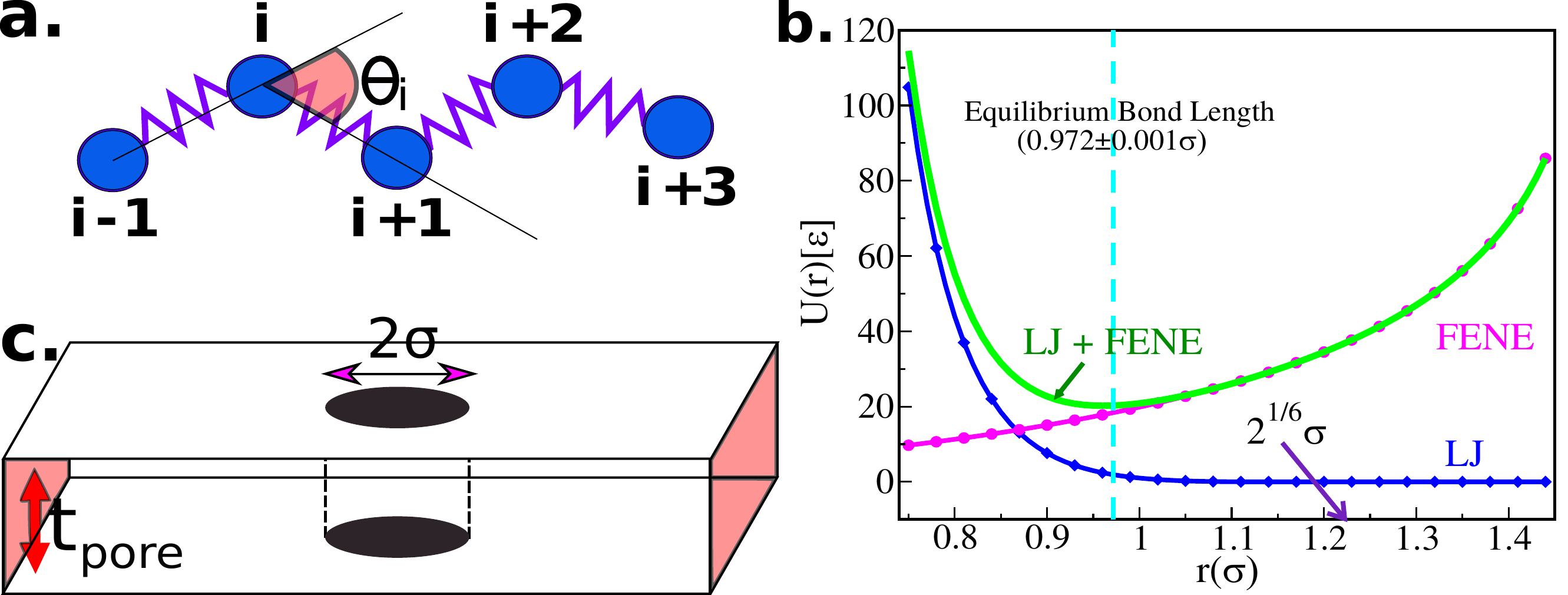}
\caption{\small \label{Interactions}(a) Illustration depicts the monomers are interacting via LJ and FENE potential. The three body bending potential is calculated using the angle $\theta_i$ between two adjacent bond vectors $\vec{b}_i$ and $\vec{b}_{i+1}$ respectively. (b) Interaction potential between two consecutive monomers is given by the green line for a separation distance $r$ in unit of $\sigma$. The blue diamonds denote the LJ potential with a cutoff radius $2^{1/6}\sigma$ and the magenta circles correspond to the FENE potential with a spring constant $\kappa_F = 30.0\epsilon/\sigma^2$. (c) A cylindrical nanopore of diameter $2\sigma$ is dilled into a material of thickness $t_{pore}$. The walls consist of purely repulsive LJ particles.}
\end{figure}

Our BD scheme is implemented on a bead-spring model of a polymer with the monomers interacting via an excluded volume (EV), a Finite Extension Nonlinear Elastic (FENE) spring potential, and a bond-bending potential enabling variation of the chain persistence length $\ell_p$ (Fig.\ref{Interactions}). The model, originally introduced for a fully flexible chain by Grest and Kremer~\cite{Grest}, has been studied quite extensively by many groups using both Monte Carlo (MC) and various molecular dynamics (MD) methods~\cite{Binder_Review}. Recently we have generalized the model for a 
semi-flexible chain and studied both equilibrium and dynamic properties~\cite{Adhikari_JCP_2013,Huang_JCP_2014,Huang_EPL_2014a} and studied compression dynamics of a model dsDNA inside a nanochannel~\cite{Polymers2016,MM2018} . The mutual EV interaction among any two monomers are given by the truncated Lennard-Jones (LJ) potential with a cut-off radius $2^{1/6}\sigma$
\begin{align}
U_{LJ} (r_{ij}) = \begin{cases}
4\epsilon \left[ \left(\frac{\sigma}{r_{ij}}\right)^{12} - \left(\frac{\sigma}{r_{ij}}\right)^{6}\right] + \epsilon, \text{ for } r < 2^{1/6} \\
0, \text{ otherwise }  
\end{cases}
\label{LJ}
\end{align}
where $\sigma$ is the effective diameter of a monomer and $\epsilon$ is the interaction strength. To mimic the connectivity between two adjacent monomers, finite-extensible-non-linear elastic (FENE) potential 
\par 
\begin{align}
U_{FENE}(r_{ij}) = - \frac{1}{2}\kappa_F R_0^2 \ln\left[ 1- \left(\frac{r_{ij}}{R_0}\right)^2\right]
\label{FENE}
\end{align}
is used with the maximum bond-stretching length $R_0 = 1.5\sigma$ and spring constant $\kappa_F=30 \epsilon/\sigma^2$. Here, $r_{ij} = |\vec{r}_{i} - \vec{r}_{j}|$ is the separation distance between two adjacent monomers $i$ and $j = i \pm 1$ located at $\vec{r}_i$ and $\vec{r}_j$ respectively. Along with these two potentials, we introduce a bending potential 
\begin{align}
U_{bend}(\theta_i) = \kappa \left( 1-\cos\left(\theta_i\right) \right) 
\label{kappa}
\end{align}
with bending rigidity $\kappa$. In three dimensions, {for $\kappa \ne 0$}, the persistence length $\ell_p$ of the chain is related to $\kappa$ via~\cite{Landau}
\begin{equation}       
\ell_p = \frac{\kappa}{k_B T},
\label{l_p}
\end{equation}
where $k_B$ is the Boltzmann constant and $T$ is the temperature.
Here $\theta_i$ is the bond angle between two subsequent bond vectors $\vec{b}_i=\vec{r}_{i+1}-\vec{r}_i$ and $\vec{b}_{i-1} = \vec{r}_i-\vec{r}_{i-1}$.
A cylindrical nanopore of diameter $2\sigma$ is drilled through a solid material of thickness $t_{pore}$ consists of immobile and purely repulsive LJ particles. Our model of DNA polymer consists $1016$ monomer beads along with $8$ heavier tags ($T_1$ - $T_8$) located at positions $154, 369, 379, 399, 614, 625, 696$, and $901$ respectively (please refer to Fig.~2 and Table-I in the main article).  A recent study by Zhang et al. on 48512 bp long dsDNA uses 75 bp long protein tags as barcodes~\cite{Reisner-Small-2018}. In simulation, we purposely choose the mass of a tag ($m_{tag}$) three times heavier of a normal monomer to replicate the tags used in the experiments. We proportionally increase the solvent friction of the tags $\Gamma_{tag} = 3 \Gamma_i$. We use the Brownian dynamics to solve the equation of motion of a monomer $i$ having a mass $m_{i}$ and solvent friction $\Gamma_{i}$ as
\begin{align}
m_{i} \ddot \vec{r_i} = \vec{\nabla_i} \left[ U_{LJ} + U_{FENE} + U_{bend} + U_{wall}\right] - \Gamma_{i} \vec{v_i} + \eta_i  
\label{INTEGRATION}
\end{align} 
where $\Gamma_{i} = 0.7\sqrt{m_{i}\epsilon^2/\sigma^2}$ is the frictional coefficient arising from solvent-monomer interaction. For the case of a tag, $m_{tag} = 3 m_i$ and $\Gamma_{tag} = 2.1\sqrt{m_{i}\epsilon^2/\sigma^2}$. The Gaussian white noise $\eta_i$ arising from thermal fluctuation is delta correlated and expressed as $\langle \eta_i(t). \eta_j{j}(t') \rangle = 2dk_BT \Gamma \delta_{ij} \delta(t-t')$ with $d=3$ in three dimension. We express length and energy in units of $\sigma$ and $\epsilon$ respectively such that $k_BT/\epsilon =1.0$. The parameters for FENE potential in Eq.~(\ref{FENE}) are $\kappa_F$ and $R_0$, and set to be $\kappa_F=30 \epsilon/\sigma^2$ and $R_0 = 1.5\sigma$. The numerical integration of Eq.~(\ref{INTEGRATION}) is implemented using the algorithm introduced by Gunsteren and Berendsen~\cite{Langevin}. Our previous experiences with BD simulation suggests that for a time step $\Delta t = 0.01$ these parameters values produce stable trajectories over a very long period of time and do not lead to unphysical crossing of a bond by a monomer~\cite{Huang_JCP_2014,Huang_EPL_2014a}. The average bond length stabilizes to $\langle b_l \rangle=0.971 \pm 0.001 \sigma$ with negligible fluctuation regardless of the chain size and rigidity~\cite{Huang_JCP_2014}. Hence we relate the polymer's contour length $L$ and the number of monomers $N$ as $L=(N-1)\langle b_l \rangle$.
\setcounter{figure}{0}
\renewcommand{\thefigure}{B\arabic{figure}}

\end{document}